\documentclass[12pt]{article}
\usepackage [paperwidth=8.5in,paperheight=11in,top=1.5in,bottom=0.5in, left=1.25in, right=1in]{geometry}
\usepackage{graphicx} 
\usepackage{epsfig}
\usepackage{float}
\usepackage{palatino,eulervm}
\usepackage[T1]{fontenc}
\usepackage{color}
\definecolor{gray87}{gray}{0.87}
\definecolor{gray64}{gray}{0.64}
\definecolor{gray50}{gray}{0.5}
\usepackage{amsmath}
\usepackage{titlesec}
\titleformat*{\section}{\large\bfseries}
\titleformat*{\subsection}{\normalsize\bfseries}
\usepackage{hyperref}

\begin{document}

\thispagestyle{plain}

\noindent
\rmfamily
\vspace{6pt}
\begin{center}
\large $h_{PI}$: The Citation Index for Principal Investigators \\
\normalsize
\vspace{36pt}
Christoph Steinbr\"{u}chel \\
Rensselaer Polytechnic Institute \\
Troy, NY 12180, U.S.A. \\
Email: steinc4@rpi.edu \\
\end{center} 
\vspace{36pt}

\begin{quotation}
\noindent
\large Abstract \\[8pt]
\normalsize
\noindent

A new citation index $h_{PI}$ for principal investigators (PIs) is defined in analogy to Hirsch's index $h$, but based on renormalized citations of a PI's papers. To this end, the authors of a paper are divided into two groups: PIs and non-PIs. A PI is defined as an assistant, associate or full professor at a university who supervises an individual research program.  The citations for each paper of a certain PI are then divided by the number of PIs among the authors of that paper. Data are presented for a sample of 48 PIs who are senior faculty members of physics and physics-related engineering departments at a private research-oriented U.S. university, using the ISI Web of Science citations database.  The main result is that individual rankings based on $h$ and $h_{PI}$ differ substantially. Also, to a good approximation across the sample of 48 PIs, one finds that $h_{PI} = h \,/ \sqrt{<N_{PI}>}$ where <$N_{PI}$> is the average number of principal investigators on the papers of a particular PI.  In addition, $h_{PI} = \frac{1}{2} \sqrt{C_{tot}\,/<N_{PI}>}$, where $C_{tot}$ is the total number of citations. Approaches to broadening the scope of $h$ or $h_{PI}$ are discussed briefly, and a new metric for highly cited papers called $h_x$ is introduced which represents the average number of citations exceeding the minimum of $h^2$ in the $h$-core.

\vspace{18pt}
\end{quotation}

\pagebreak
\thispagestyle{plain}
\section{Introduction}
\numberwithin{equation}{section}
\numberwithin{figure}{section}
Publication of results in research papers is a primary outcome of a research program, and the impact of such work is reflected most directly in the citations it has received.  Citations can now be analyzed easily thanks to the publication of citation indexes such as the ISI Web of Science (Thomson-Reuters, now Clarivate Analytics) and Scopus (Elsevier). These indexes are  databases compiled from references in papers published in scientific journals and possibly other types of publications (e.g. conference proceedings, books, patents, etc.). A citation index is also available through Google Scholar. Each of these indexes is capable of providing, among other statistical information, a list of papers and their citations for a specific author on those papers. Of course, different indexes are not directly comparable, as the methods of identifying citations and the source publications they cover may be different.

The use of citations was advanced greatly by a paper by Hirsch [1], who introduced a new metric called h-index which is defined as follows:

\vspace{18pt}
\begin{center}
\begin{minipage}[h]{0.8\linewidth}
\noindent
A scientist with a total number of papers $N_p$ has an h-index of $h$ if $h$ of his or her papers have at least $h$ citations each and the other $(N_p - h)$ papers have $\leq h$ citations each.
\end{minipage}
\end{center}
\vspace{18pt}
The h-index is visualized readily with help of a graph of citations $C(r)$ of papers labeled $r$, where the papers are listed in descending order of the number of citations received. The h-index is regarded as an indicator of the scientific impact of the work of a researcher rather than just a straightforward measure of research productivity such as $N_p$ or the total number of citations received, $C_{tot}$. Hirsch also noted that, by definition, for a citation record $C(r)$ with h-index $h$, the minimum number of total citations $C_{tot}$ is $h^2$. Furthermore, Hirsch suggested an empirical relation
\begin{equation}
   C_{tot} = ah^2
\label{eq01}
\end{equation}

\noindent
where $a$ is a constant with a value between 3 and 5, but no data were provided.

The h-index was recognized quickly as a major advance in making use of citation data [2,3]. At the same time it was also criticized as being limited on the basis of four types of argument. First, it was argued that the definition of $h$ was somewhat arbitrary, in the sense that one could have defined a similar, but more general index $h_q$ as given by the number of papers $h_q$ having received at least $q\times h_q$ citations [4,21].  So the original $h$ is just a special case of a class of possible indexes $h_q$, namely the one with $q = 1$.  (Thus $h$ should perhaps be called $h_1$). It is not immediately apparent why $h_1$ should be preferable over other similar indexes $h_q$.

The second argument concerned the fact that $h$ was just a single number and thus could not describe all important aspects of the shape of the distribution {C(r)}. For example, $h$ is insensitive to additional citations of papers already in the $h$-core, i.e. papers with citations $>h$, especially very highly cited papers. In order to remedy this situation, other indices have been introduced emphasizing highly cited papers. Egghe proposed his g-index $g$ as the highest number $g$ of papers that together received $g^2$ or more citations [5]. $g$ is always higher than the corresponding $h$, and $g$ and $h$ have been compared extensively  [6,7]. 
Zhang introduced his e-index as a direct measure of the citations exceeding those in the $h$-core [8]. Jin focused on the total number of citations in the $h$-core [9]. A disadvantage of $g$ and $e$ is that their relationship to the shape of {C(r)} is not as evident as it is for $h$. It has also been suggested to use additional values on the {C(r)} curve, such as {C(0.5h)} or {C(1.5h)} [10], or {C(10h)} [11], but it is unclear whether these facilitate meaningful distinctions in practice, say for distinguishing between citation records with the same $h$. Over time, at least 20 variations of the h-index have been proposed, and the extent to which they are independent and provide additional information has been examined [2,3,12,21].

The third argument suggests that the h-index as a single number for ranking scientists exhibits  a certain counterintuitive behavior in that it does not always produce the expected result that if two scientists achieve the same relative performance improvement, their ranking relative to each other should remain unchanged. This finding is related to the fact that $h$, being a size-dependent quantity, never goes down as $C_{tot}$ increases [13].

The fourth argument focuses on the fact that citations are to papers, but papers are often  authored by several people. A citation can only be associated uniquely with an author for a single-author paper. There is no clear a priori justification why each author of a multi-author paper should claim credit to all citations of that paper. This multi-author issue has been addressed by dividing citations among the authors of a paper in various ways [6,14-19,21], e.g. by assigning each author an equal fraction of the citations, or by assigning authors different weights in parsing the citations depending on a perceived rank order among those authors. A modified h-index is then determined from the recalibrated citations. However, there is no consensus about how to do this in a universally acceptable manner.

The present paper seeks to make two points. The first one is to deal with part of the multi-author problem by introducing a \textbf{new index $h_{PI}$}, the \textbf{\textit{h-index for principal investigators}}. $h_{PI}$ can be seen as new type of coauthorship-weighted index [21]. To this end, the authors of a paper are divided into two groups, principal investigators (PIs) and non-PIs. (The method of how this was done is described in detail in the Methodology section below). The assumption is that, for the purpose of comparing different PIs and their research programs, it is appropriate to \textbf{\textit{assign equal fractional credit for a citation to each PI}}. The $h_{PI}$ for a particular PI is constructed like $h$, but from citations for each paper divided by the number of PIs among the paper's authors. Citation data for physics and engineering faculty at a private, research-oriented technical U.S. university are used for illustration purposes. The main result is that rankings based on $h_{PI}$ may differ markedly from rankings based on $h$. Other results include various relationships between $h_{PI}$, $h$, and the average number of PIs, <$N_{PI}$>, on a particular PI's papers. 

The second point of the present paper is to re-examine approaches to dealing with highly cited papers and to propose a new metric for assessing such papers.

\section{Methodology}
\numberwithin{equation}{section}
\numberwithin{figure}{section}
For the purpose of the present study, a citation will be the listing of a certain scientific paper as a reference in a subsequent scientific paper, as reported in the ISI Web of Science Core Collection database (referred to subsequently as WoS).

A principle investigator (PI) is defined as an assistant, associate, or full professor at an academic institution. Such an individual is expected to be responsible for planning and executing an identifiable, independent academic research program, including raising funds to do so.

The sample of PIs studied here consisted of 48 senior faculty members at the Rensselaer Polytechnic Institute in Troy, NY, U.S.A. They are all tenured associate of full professors in the Physics department and various Engineering departments (a few have emeritus status). There is no doubt that each one of them would agree that he or she, as well as everyone else among the 48 individuals, is indeed a PI. In addition, they are all known to me personally as PIs running, or having run, their own research program. Their research can be categorized as small-group and based in condensed-matter science.  

The WoS citation records of the 48 individuals were examined as of the beginning of March 2017 and extending back to 1970. For each PI, the number of citations to each paper was divided by the number of PIs on the paper. If there were authors who were not affiliated with a university, e.g. members of a national or an industrial laboratory, these were counted as PIs. For the papers where the current PI earlier had been a Ph.D. student, postdoctoral fellow, or assistant professor, he or she was also counted as a PI on those papers. In the small number of papers where a current assistant professor was among a paper's authors together with associate or full professors, the assistant professor was counted as a PI for the purpose of ascertaining the number of PIs on the paper. The $h_{PI}$ index was determined from the recalibrated citation counts. 

In addition, a similar index $h_{A}$ for authors was compiled from citations recalibrated on a per author basis, for occasional comparison with $h_{PI}$ data. Other parameters such as the average number of PIs, <$N_{PI}$>, and the average number of authors, <$N_A$>, for the papers of each individual PI were also calculated. 

Search results from the WoS were reviewed carefully for each of the 48 PIs to insure that only papers from the PI in question were counted. This turned out to be non-trivial. For example, in one case the last name and first initial of a PI were identical for two individuals at the same university. In this case, the WoS search returned a combined list of citations, which had to be corrected manually.

\section{Results}
\numberwithin{equation}{section}
\numberwithin{figure}{section}
\subsection{Comparison of $h$ and $h_{PI}$}

The range of numerical values for the relevant citation parameters across the population of the 48 PIs is summarized in the table below.  

\begin{center}
\textbf{Table 3.1: Characteristics of the sample of 48 PIs}
\end{center}

\begin{center}
\renewcommand{\arraystretch}{1.5}
 \begin{tabular}[!htbp]{| l | l |}
  \hline 
  \hspace{2pt} $C_{max}$ \hspace{2pt}&\hspace{2pt} 22 -- 1803 \\
  \hspace{2pt} $C_{tot}$ \hspace{2pt}&\hspace{2pt} 196 -- 17876 \hspace{2pt} \\ 
  \hspace{2pt} <$N_{A}$> \hspace{2pt}&\hspace{2pt} 1.89 -- 7.14 \\
  \hspace{2pt} <$N_{PI}$> \hspace{2pt}&\hspace{2pt} 1.22 -- 4.14 \\
  \hspace{2pt} $h$  \hspace{2pt}&\hspace{2pt} 5 -- 68 \\
  \hspace{2pt} $h_{PI}$  \hspace{2pt}&\hspace{3pt} 4 -- 58 \\
  \hline    		       
 \end{tabular}
\end{center}

\vspace{15pt}
$C_{max}$ refers to the range of maximum citations received across the sample of 48 PIs, and $C_{tot}$ to the total number of citations. For each PI, the averages <$N_{PI}$> and <$N_A$> were computed up to and including the paper at {h}. Including more papers did not alter these numbers significantly. The grand total number of papers included for all PIs together was 6629, and the grand total of all citations was 164414, but there is some double-counting involved in these numbers due to collaborations between PIs in the sample.

It should also be understood that the ranges in the table are taken across the entire population of 48 PIs, so that e.g. the PI with the highest number of total citations $C_{tot}$, i.e. 17876, is not necessarily the same as the PI with the highest $C_{max}$ of 1803.

Another point about the set of 48 PIs is that, on average, there were about 2 authors per PI. That is, on average each PI collaborated with about one non-PI on a project. This again illustrates the small-group nature of the research efforts examined here. However, on an individual basis, <$N_{A}$>/<$N_{PI}$> varied between about 1.5 and 3, which implies that the average number of non-PIs per PI varied between about 0.5 and 2.

\begin{figure}[htb]
    \begin{minipage}[h]{12pt}           
      \hspace{12pt}			
    \end{minipage}
    \begin{minipage}[h]{0.7\linewidth}
      \epsfig{file=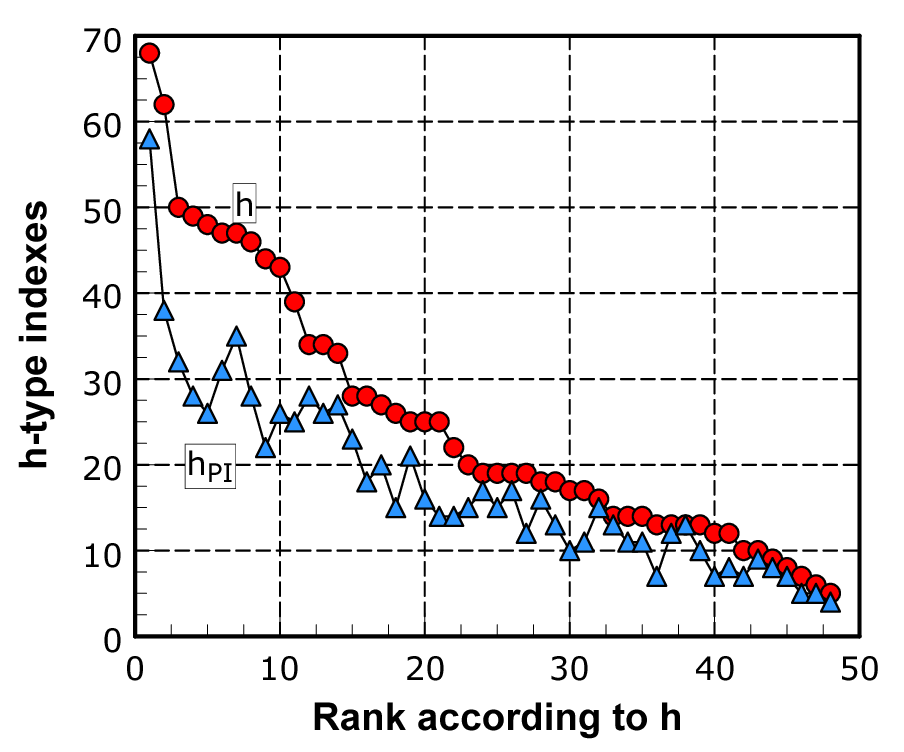,width=10cm}
    \end{minipage} \hspace{6pt}	
    \begin{minipage}[h]{0.2\linewidth}
      \caption{\, \hspace{24pt}} Ranked display of all 48 PIs by $h$, and comparison to $h_{PI}$.
      \label{h-hPI2}
    \end{minipage}
\end{figure}

An overview of the relationship between $h$ and $h_{PI}$ is presented in Fig.~\ref{h-hPI2} where $h$ and $h_{PI}$ are displayed in a ranked fashion, i.e in order of decreasing $h$. You will notice first that $h$ > $h_{PI}$. This is, of course, because $h$ = $h_{PI}$ would only be true if a particular PI were the only PI on all of his or her papers. For many PIs, in particular at higher $h$, one finds that $h_{PI}$ is much smaller than $h$, .   

The conclusion from this graph is that a ranking on the basis of $h_{PI}$ would be significantly different than the ranking based on $h$. For example, the PI ranked 5th with $h$ = 48 would be ranked 11th or 12th based on an $h_{PI}$ of 26. In addition, the range of values for $h_{PI}$ is compressed substantially compared to $h$, especially at the high-$h$ end, because PIs with higher $h$ tend to cooperate with more PIs. The exception is PI No. 1 with the highest $h$ of 68 and only a slightly reduced $h_{PI}$ of 58, thanks to a fairly low <$N_{PI}$>.

\subsection{Relationships between ${h}$, ${h_{PI}}$ and <${N_{PI}}$>}

\begin{figure}[h!tb]
\centering
\includegraphics [width=13cm] {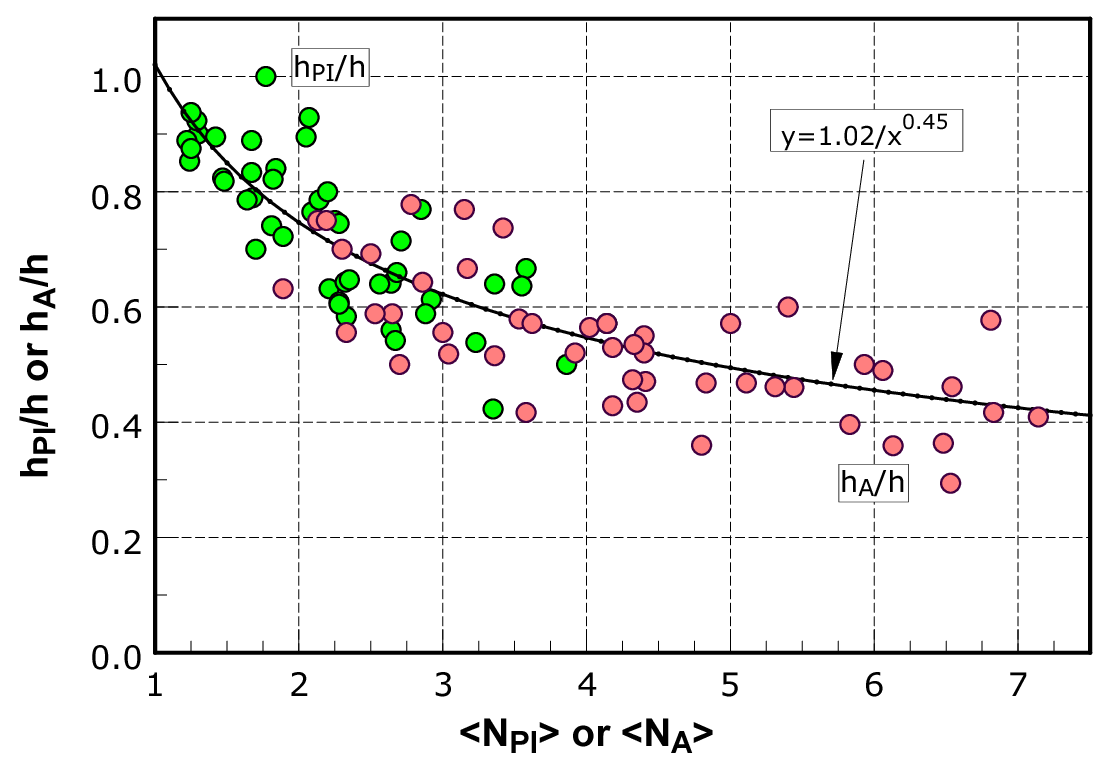}
\caption{\, Plot of ${h_{PI}}/{h}$ vs. <${N_{PI}}$> and ${h_{A}}/{h}$ vs. <${N_{A}}$>. The x-scale refers to either the average number of PIs or the average number of authors.}
\label{hPI/h-NPI}
\vspace{9pt}
\end{figure}

First, it turns out that there is no simple correlation between ${h}$ and <${N_{PI}}$>. This is consistent with Fig.~\ref{h-hPI2} since the relative deviations of ${h_{PI}}$ from ${h}$, which are due to variations in <${N_{PI}}$>, are distributed in a rather irregular fashion. However, some interesting conclusions can be drawn from a plot of ${h_{PI}}/{h}$ vs. <${N_{PI}}$> (Fig.~\ref{hPI/h-NPI}).

\vspace{6pt}
Fig.~\ref{hPI/h-NPI} shows ${h_{PI}}/{h}$ vs. <${N_{PI}}$> and ${h_{A}}/{h}$ vs. <${N_{A}}$> on the same x-scale. The solid curve is a best fit to the ${h_{PI}}/{h}$ data and has the form $y=1.02/x^{0.45}$. The same curve also describes the ${h_{A}}/{h}$ data quite well although in this latter case the best fit has the form $y=0.95/x^{0.42}$. In summary, the data for ${h_{PI}}$ vs. <${N_{PI}}$> can be described reasonably well by a simple expression of the form
\begin{equation}
   h_{PI} = \frac{h}{\sqrt{<N_{PI}>}}
\label{eq02}
\end{equation}

\vspace{9pt}
\noindent
and a corresponding equation is valid between ${h_A}$, $h$, and <$N_{A}$>. Again, it should be emphasized that Eq.~\ref{eq02} applies in an average sense, and as Fig.~\ref{hPI/h-NPI} shows, there is substantial scatter of the data about the fitted mathematical curve.

Next, we turn our attention to the relationship between the indexes ${h}$ and ${h_{PI}}$ and the total number of citations $C_{tot}$. With a view on Eqs.~\ref{eq01} and ~\ref{eq02}, we plot $\sqrt{C_{tot}}$ vs. $h$ and $\sqrt{(C_{tot}/<N_{PI}>)}$ vs. $h_{PI}$ (see figure below).

\pagebreak

\begin{figure}[hb]
\centering
\includegraphics [width=8.0cm] {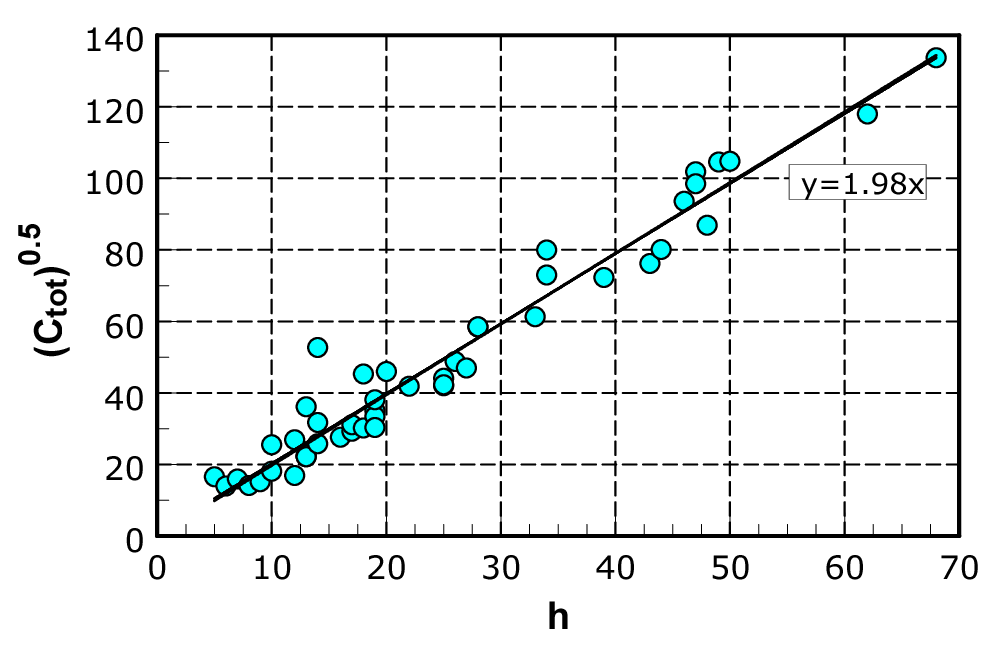} 
\vspace{3pt} \includegraphics [width=7.3cm] {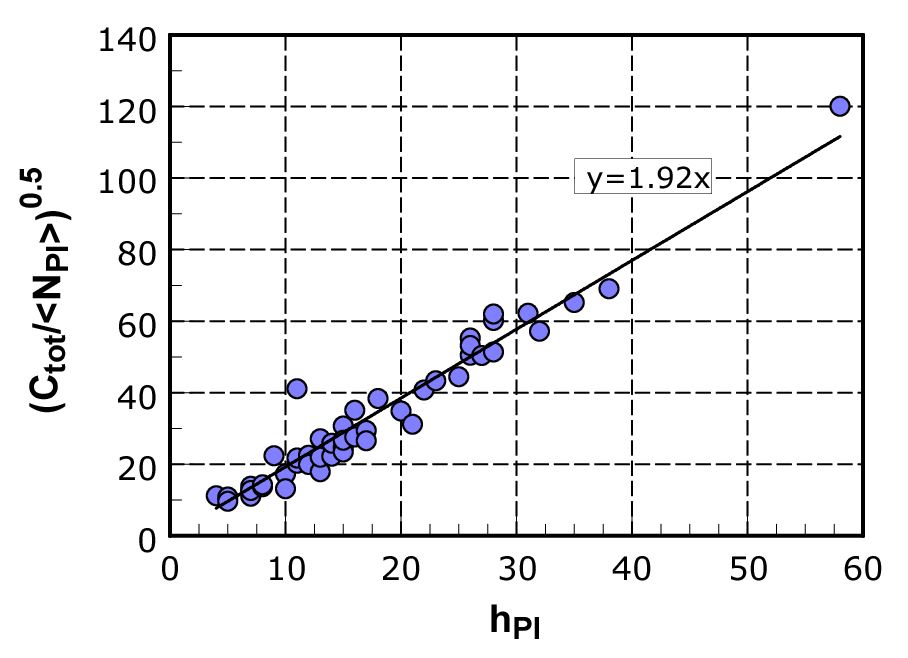}
\caption{Plots of $\sqrt{C_{tot}}$ vs. $h$ and $\sqrt{(C_{tot}/<N_{PI}>)}$ vs. $h_{PI}$ showing the indexes as proportional to the square root of the normalized total citations.}
\label{Ctot-h-hPI}
\end{figure}

It is apparent that the two linear fits are excellent. Thus, to a good approximation, the relationships between the two indexes and $C_{tot}$ can be written as follows:

\vspace{6pt}
\begin{equation}
   h = \frac{1}{2}\, \sqrt{C_{tot}}
\label{eq03}
\end{equation}

\vspace{6pt}
\begin{equation}
   h_{PI} = \frac{1}{2}\, \sqrt{\frac{C_{tot}}{<N_{PI}>}}
\label{eq04}
\end{equation}

\vspace{15pt}
\noindent
Equivalently, to a good approximation,

\begin{equation}
   C_{tot} = 4 h^2
\label{eq05}
\end{equation}

\begin{equation}
   C_{tot}\,/<N_{PI}>\,\, = 4 h_{PI}^2
\label{eq06}
\end{equation}

\vspace{12pt}
\noindent
Hence, the constant $a$ in Hirsch's equation \ref{eq01}, $C_{tot} = a h^2$, is about equal to 4, and the same type of equation applies to $h_{PI}$ if $C_{tot}$ is replaced by $C_{tot}$/<$N_{PI}$>. The relationship ~\ref{eq05} observed here is in agreement with Redner's results [23].

Note also that the data of Fig.~\ref{Ctot-h-hPI} show considerable scatter about the straight lines, especially at lower values of $h$. Thus Eqs.~\ref{eq03} and \ref{eq04} should not be used as accurate predictors of $h$ and $h_{PI}$.

\vspace{3pt}
\subsection{Going beyond $h$ and $h_{PI}$}

A major question in using citation indexes is to what extent $h$ or $h_{PI}$ provide a proper metric for distinguishing between individual achievements. In particular, how can one arrive at a meaningful distinction between citation records having the same $h$? Clearly, this will require more than just describing an entire distribution $C(r)$ by a single number. The real question is how to broaden the scope of $h$ and $h_{PI}$, so as to capture accurate additional information regarding high citations, to which $h$ is not sensitive.

A first idea is to use $h$ together with another of the $h_q$ indexes, for example with $h_2$ or $h_{0.5}$, depending on which part of $C(r)$ one is interested in [10]. An illustration is presented in Fig.~\ref{CScits}:

\begin{figure}[htb]
    \begin{minipage}[h]{0.4\linewidth}
      \epsfig{file=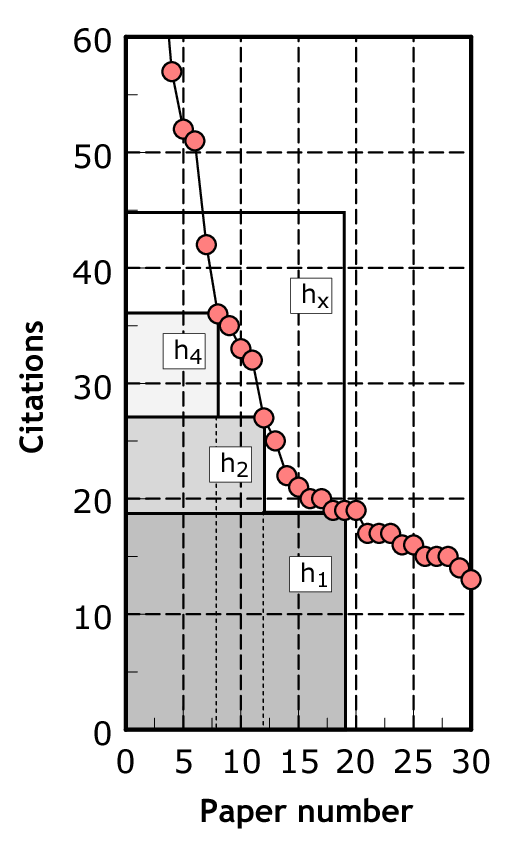,width=6cm}
    \end{minipage} \hspace{30pt}	
    \begin{minipage}[h]{0.45\linewidth}
\caption{\,Sample $C(r)$ for $h_1$ = 19, and comparison with $h_2$, $h_4$, and $h_x$. Papers ranked 1-3 are off the chart.}
      \label{CScits}
\vspace{12pt}
This figure displays part of a $C(r)$ plot for which $h = h_1 = 19$. Also shown are $h_2 = 12$, $h_4 = 8$, and $h_x = 26$. ($h_x$ will be defined below in Eq.~\ref{hx}). Note that the index $q$ in $h_q$ can be interpreted as the aspect ratio of the rectangle inscribed into $C(r)$ at $h_q$ as long as the $x$- and $y$-axes have the same units. Here the issue is whether, for a given  $h_1$, parameters such as $h_2$ and $h_4$, or perhaps $h_8$ or $h_{10}$, will discriminate adequately between different curves $C(r)$ with respect to highly cited papers. 
    \end{minipage}
\end{figure}

On the other hand, it may be advantageous to use a suitable integral property of $C(r)$, i.e. a property involving a sum of citations, rather than just another point on $C(r)$. Several possibilities come to mind:

\begin{enumerate}
  \item {} One could use the sum of all citations in the $h$-core, i.e. the sum of all citations in $C(r)$ down to the paper with rank $h$. Let this quantity be named $C_h$:
  \begin{equation}
     C_h = \sum_{r=1}^{h} C(r)
  \label{Ch1}
  \end{equation}

This is a direct measure of the first part of $C(r)$, i.e. the more highly cited papers (see [9]).

  \item {} Since an individual with a certain h-index $h$ must have at least $h^2$ citations, it might allow for more precision to use the sum of the citations in the $h$-core in excess of $h^2$. Let this quantity be named $C_{h,x}$:
  \begin{equation}
     C_{h,x} = \sum_{r=1}^{h} C(r) \, - \,h^2
  \label{Chx1}
  \end{equation}

$C_{h,x}$ is the basis for Zhang's e-index $e$, defined as the square root of $C_{h,x}$ [8]:
  \begin{equation}
     e = \sqrt{C_{h,x}}
  \label{he}
  \end{equation}

  \item {} An index $h_x$ for citations in excess of $h$ can be derived from $C_{h,x}$ above by dividing by $h$:
  \begin{equation}
     h_x = C_{h,x}\, / \,h = \frac{1}{h}\,\sum_{r=1}^{h} C(r) \,-\,h
  \label{hx}
  \end{equation}

$h_x$ has a simple geometric interpretation with help of the $C(r)$ curve: $h_x$ is the average of the citations in excess of the $h$-core and can be visualized as the height of the rectangle on top of the $h^2$ square in Fig.~\ref{CScits} above.
\end{enumerate} 

\vspace{6pt}
Which of all these possibilities may be useful in practice will be examined with a select set of 8 comparable PIs from the whole sample of 48 PIs. These are the 8 PIs in Fig.~\ref{h-hPI2} with $h$ values from 50 to 43. The relevant data are displayed in the next figure.

\begin{figure}[htb]
\centering
\includegraphics [width=11cm] {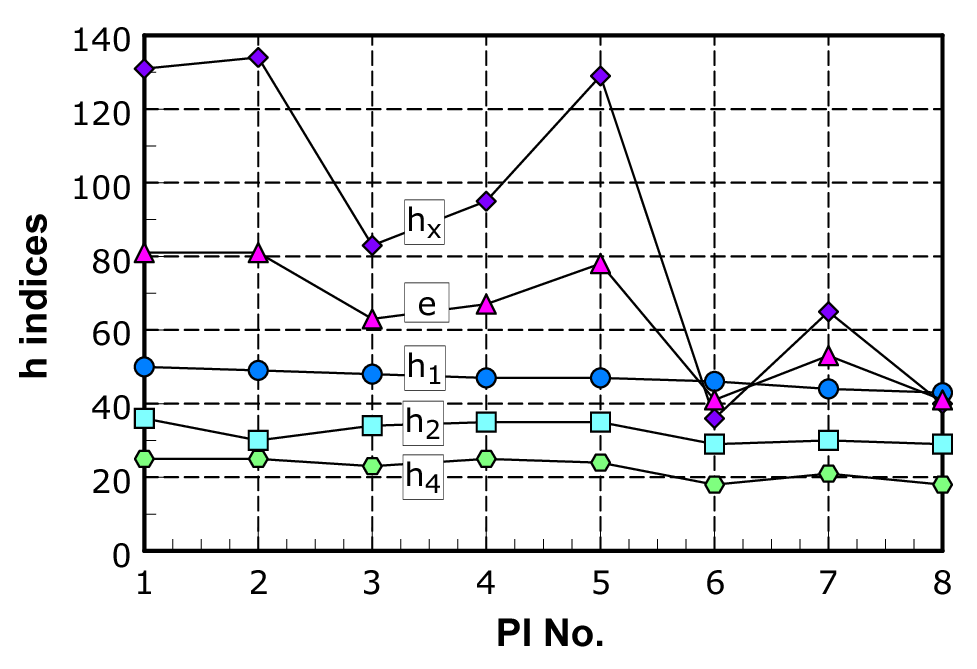}
\caption{\, Data for 8 PIs showing $h$-related indices as indicators of excess citations relative to the $h$-core. Note the large variations in $h_x$ and $e$ for records with similar $h_1$.}
\label{h12x}
\end{figure}

\vspace{6pt}
Fig~\ref{h12x} displays $h_1$ = $h$, $h_2$, $h_4$, $e$ (see Eq.~\ref{he}), and $h_x$ (see Eq.~\ref{hx}). Focusing first on $h_2$ and $h_4$, if those indexes were indeed good indicators of more papers with high citations, then for the PIs No. 5 to 8, we might expect PI No. 5 to have the most highly cited papers and PIs No. 6 and 8 the least. On the other hand, for the PIs No. 1 to 5, $h_2$ and $h_4$ present an inconsistent picture with respect to relative differences. 

The picture is clarified by $e$ and $h_x$ in Fig.~\ref{h12x}. Both these indicators clearly  show the differences in the citations in excess of the total core minimum of $h^2$, but $h_x$ does so more prominently than $e$. From both $h_x$ and $e$ it is evident that PIs No. 1, 2, and 5 have the highest excess citations and PIs No. 6 and 8 the lowest.  However, $h_x$ seems more transparent to interpret in relation to $C(r)$ since it directly represents the average of the core citations in excess of $h$. Keep in mind that the average number of citations in the entire core is given by $h + h_x$ (see Eq.~\ref{hx}). Therefore, one can deduce from $h_x$, for example, that PI No. 7 has about 50\% more core excess citations than PI No. 8 ($h_x \approx$ 60 vs 40) and about 25\% more total core citations ($h+h_x \approx$ 100 vs 80).

The results from Fig.~\ref{h12x} can be summarized as follows: The highly cited papers in the $h$-core are visualized most easily by comparing $h$ and $h_x$. The indexes $h_2$ and $h_4$ are much less sensitive and are not reliable indicators of highly cited papers. Thus it appears likely that indices $h_q$ with $q$>1 in general will not provide a sufficiently accurate description of the high-citation end of the citation distribution $C(r)$.

Another conclusion from Fig.~\ref{h12x} is that if a ranking of authors were to be done on the basis of, say, $h_2$ or $h_4$, such a ranking would be different from the usual one based on $h_1$ (i.e. on $h$). This agrees with a conclusion reached previously by Schreiber [22].

\section{Discussion and Conclusions}

The main points of this paper have been: 1) to introduce the citation index $h_{PI}$ for principal investigators, defined here as Assistant, Associate and Full Professors at an academic institution, and 2) to propose the index $h_x$ as a transparent measure of highly cited papers in the $h$-core.

The new index $h_{PI}$ is constructed much like the conventional Hirsch index $h$, but from citations renormalized per PI.  That is, citations for each paper on which a particular researcher was one of the PIs are divided by the number of PIs on the paper. This was to resolve at least in part the multi-author dilemma, in this case to account more accurately for the impact of an individual PI who often publishes with other PIs. 

The main result is that $h_{PI}$ is always lower, often considerably lower, than $h$. Across the  sample of 48 PIs, $h_{PI}$ scales as $h/\sqrt{<N_{PI}>}$, where <$N_{PI}$> is the average number of PIs on the papers of a particular PI. This means that a PI with <$N_{PI}$> = 4 can be expected to have a $h_{PI}$ reduced by a factor 2 compared to $h$. In addition, $h$ scales as $\sqrt{C_{tot}}$ and $h_{PI}$ as $\sqrt{C_{tot}/<N_{PI}>}$, where $C_{tot}$ is the total number of citations.  

It seems evident that if one is interested in comparing different researchers, the impact of a particular PI's work as an individual, supervising a research program, is reflected better by $h_{PI}$ than by $h$. It would appear difficult to argue that a PI should claim full credit for the citations of a paper on which he or she is one of several, or indeed many, PIs. At the same time, it seems fair to apportion credit equally among all PIs concerned, for the purpose of comparing PIs. It also follows that relative rankings of PIs based on their $h$ should not be taken literally unless their <$N_{PI}$> are very similar. In any event, $h_{PI}$ will provide complimentary information on the citation record of an individual with a given $h$.

Additionally, it is worth noting that the relationship in Eq.~\ref{eq05} derived from Fig.~\ref{Ctot-h-hPI} is in agreement with an earlier observation by Redner [23], who found that, for a sample of  255 physicists, the ratio of $C_{tot}/h^2$ peaked around a value of 4.

It is also clear that $h$ and $h_{PI}$ still are only single numbers and thus give a rather incomplete description of the impact of a PI's work. However, additional information can be gleaned from a more in-depth examination of the ranked distribution of citations $C(r)$, especially with respect to highly cited papers.

First, the present results indicate that any other single number, such as e.g. $h_2$ or $h_4$ or the $w$-index [11] (equivalently, $h_{10}$), does not provide a more meaningful metric for the impact of an individual's work than does $h$. Moreover, even a combination of two numbers, such as $h$ and $h_2$, or $h$ and $h_4$, does not discriminate reliably between different citation records with respect to highly cited papers. The same conclusion applies to the suggestion to use $h$ in combination with the number of citations of the most highly cited paper, unless one is of the opinion that one especially highly cited paper is indicative of special merit [20].

Valuable information about highly cited papers can be derived from the excess citations in the $h$-core, $C_{h,x}$ (see Eq.~\ref{hx}), i.e. the citations above $h^2$ (or, citations larger than $h$). The most transparent visualization of $C_{h,x}$ is by way of the index $h_x$, which is defined as $h_x = C_{h,x}/h$. That is, $h_x$ represents the average of the citations exceeding $h$ (Fig.~\ref{CScits}), so that $h_x$ + $h$ = $C_h$ is the average of all the citations of the papers in the $h$-core (Eq.~\ref{hx}). Thus a more accurate assessment of a citation record can be achieved by combining $h$, or better yet $h$ and $h_{PI}$, with an integral measure such as $h_x$.

Of course, with all these citation-based indicators the fundamental issue is not so much what they measure, and with what precision, but rather the judgment of how these metrics should be valued. For example, is the most highly cited paper most important, and if so why? Does it  really reflect research achievement, or is it due to a review article often pointed to as background for the research field in question? 

Perhaps the most important value question is:  What differences in $h$, or for that matter $h_{PI}$, indicate significant differences in research achievement? Hirsch initially suggested that differences of about 10 in $h$ would do so [1]. In his 2010 paper he was more specific, saying that ".. in physics reasonable values of $h$ (with large error bars) might be $h \sim$ 12 for advancement to tenure (associate professor), $h \sim$ 18 for advancement to full professor, $h \sim$ 20-25 for fellowship in the American Physical Society" [17]. In light of the present results, this judgment ought to be re-examined. Perhaps it should refer to values for $h_{PI}$ rather than $h$, because $h_{PI}$ reflects more accurately the achievements of an individual. Also note that the range in $h_{PI}$ among a set of individuals will generally be much reduced in comparison to the range in $h$.

If the judgment were that all authors should indeed receive equal credit, then an author-based h-index $h_A$ can be obtained along the same lines as described for $h_{PI}$, by dividing the number of citations of each paper by the number of its authors [15]. In that case, I have found  that, on average, $h_A = h/\sqrt{<N_A>}$ where <$h_A$> is the average number of authors. This means that using $h$/<$N_A$> to account for multiple authors [6], i.e. to adjust citations at the ranking level, is not justified (see [15] for a more extended discussion of this point). 

A practical consequence of the present results is that if you are up for an academic appointment or a promotion and the committee evaluating you is looking at citations, you should make them   aware of the $h_{PI}$ index when they compare you with other candidates. Comparisons based on the conventional index $h$ may be misleading. It should not be difficult for you to compile relevant citations for obtaining $h_{PI}$ since you will know the PIs in your field. If no information on the number of PIs is readily available, you should at least be able to determine the average number of authors. Then a reasonable approximation will be <$N_{PI}$> = 0.5 <$N_{A}$> for the type of small-group research of interest here, and Eq.~\ref{eq04} will provide a useful approximation for $h_{PI}$. You may also want to mention that highly cited papers can be taken into account by analyzing the excess citations in the $h$-core, preferably in the form of the   corresponding index $h_x$.

\section{References}

\vspace{3pt}
\noindent
[1] {} Hirsch J.E. (2005). An index to quantify an individual's scientific research output. 
 \emph{Proc. Nat. Ac. Sci.} \textbf{102}: 16569-16572.

\noindent
[2] {} Alonso S., Cabrerizo F.J., Herrera-Viedma E., Herrera F. (2009). h-Index: A review focused in its variants, computation and standardization for different scientific fields. \emph{J. Informetrics} \textbf{3}: 273-289.

\noindent
[3] {} Ruscio J., Seaman F., D'Oriano C., Stremlo E., Mahalchik K. (2012). Measuring Scholarly Impact Using Modern Citation-Based Indices. \emph{Measurement} \textbf{10}: 123-146.

\noindent
[4] {} Eck, N.V., Waltman, L. (2008). Generalizing the h- and g-indices. \emph{J. Informetrics}  \textbf{2}: 263-271.

\noindent
[5] {} Egghe L. (2006). Theory and practise of the g-index. \emph{Scientometrics} \textbf{69}:  131-152.

\noindent
[6] {} Egghe L. (2008). Mathematical Theory of the h- and g-Index in Case of Fractional Counting of Authorship. \emph{J. Am. Soc. Inf. Sci. Technol}. \textbf{59}: 1608-1616.

\noindent
[7] {} Schreiber M. (2008). An empirical investigation of the g-index for 26 physicists in comparison with the h-index, the A-index, and the R-index. \emph{J. Am. Soc. Inf. Sci. Technol}. \textbf{59}:  1513-1522.

\noindent
[8] {} Zhang C-T. (2009). The e-Index, Complementing the h-Index for Excess Citations. \\ 
\url{http://journals.plos.org/plosone/article?id=10.1371/journal.pone.0005429}

\noindent
[9] {} Jin, B. H. (2006). h-Index: An evaluation indicator proposed by scientist. \emph{Science Focus} \textbf{1(1)}: 8-9. (In Chinese)

\noindent
[10] {} Dorta-Gonzalez P., Dorta-Gonzalez M-I. (2011). Central indexes to the citation distribution: A complement to the h-index. \emph{Scientometrics} \textbf{88}: 729-745.

\noindent
[11] {} Wu Q. (2010).  The w-index: A measure to assess scientific impact by focusing on widely cited papers.  \emph{J. Am. Soc. Inf. Sci. Technol.} \textbf{61}: 609-614.

\noindent
[12] {} Schreiber M. (2010). Twenty Hirsch index variants and other indicators giving more or less preference to highly cited papers. \\
\url{https://arxiv.org/abs/1005.5227}

\noindent
[13] {} Waltman L., van Eck N.J. (2011). The inconsistency of the h-index.  \emph{J. Am. Soc. Inf. Sci. Technol}. \textbf{63}: 406-415.

\noindent
[14] {} Batista P.D., Campiteli M.G., Kinouchi O., Martinez A.S. (2006).  Is it possible to compare researchers with different scientific interests?  \emph{Scientometrics} \textbf{68}: 179-189.

\noindent
[15] {} Schreiber M. (2009).  A Case Study of the Modified Hirsch Index $h_m$ Accounting for Multiple Coauthors. \emph{J. Am. Soc. Inf. Sci. Technol}. \textbf{60}: 1274-1282

\noindent
[16] {} Zhang C-T. (2009).  A proposal for calculating weighted citations based on author rank.  \emph{EMBO Reports} \textbf{10}: 416-417.

\noindent
[17] {} Hirsch J.E. (2010).  An index to quantify an individual's scientific research output that takes into account the effect of multiple coauthorship.  \emph{Scientometrics} \textbf{85}:  741-754.

\noindent
[18] {} Tol R.S.J. (2011). Credit where credit's due: Accounting for co-authorship in citation counts. \emph{Scientometrics} \textbf{89}: 291-299.

\noindent
[19] {} Galam S. (2011). Tailor based allocations for multiple authorship: a fractional gh-index. \emph{Scientometrics} \textbf{89}: 365-379.

\noindent
[20] {} Dorogovtsev S.N. and Mendes J.F.F. (2015). Ranking scientists. \emph{NATURE PHYSICS} \textbf{11}: 882-883.

\noindent
[21] {} Todeschini R. and Baccini A. (2016). \emph{Handbook of Bibliometric Indices : Quantitative Tools for Studying and Evaluating Research} (John Wiley) pp 54-73.

\noindent
[22] {} Schreiber M. (2013). A case study of the arbitrariness of the h-index and the highly-cited-publications indicator. \emph{J. Informetrics} \textbf{7}: 379-387

\noindent
[23] {} Redner, S. (2010). On the meaning of the h-index. \emph{J. Stat. Mech.} L03005.

\end{document}